# Three Dimensional Annihilation Imaging of Antiprotons in a Penning Trap[1]

M.C. Fujiwara[ab2], M. Amoretti[c], G. Bonomi[d], A. Bouchta[d], P.D. Bowe[e], C. Carraro[cf] C.L. Cesar[g], M. Charlton[h], M. Doser[d], V. Filippini[i], A. Fontana[j], R. Funakoshi[a], P. Genova[j], J.S. Hangst[f], R.S. Hayano[a], L.V. Jørgensen[h], V. Lagomarsino[cf], R. Landua[d], D. Lindelöf[k], E. Lodi Rizzini[l], M. Macri[c], N. Madsen[e], M. Marchesotti[i], P. Montagna[ij], H. Pruys[k], C. Regenfus[k], P. Rielder[d], A. Rotondi[ij], G. Testera[c], A. Variola[c], D.P. van der Werf[h]

(ATHENA Collaboration)

[a] *Department of Physics, University of Tokyo, Tokyo 113-0033 Japan*
[b] *Atomic Physics Laboratory, RIKEN, Saitama, 351-0189 Japan*
[c] *Istituto Nazionale di Fisica Nucleare, Sezione di Genova, 16146 Genova, Italy*
[d] *EP Division, CERN, Geneva 23 Switzerland*
[e] *Department of Physics and Astronomy, University of Aarhus, DK-8000 Aarhus, Denmark*
[f] *Dipartimento di Fisica, Università di Genova, 16146 Genova, Italy*
[g] *Instituto de Fisica, Universidade Federal do Rio de Janeiro, Rio de Janeiro 21945-970, Brazil*
[h] *Department of Physics, University of Wales Swansea, Swansea SA2 8PP, UK*
[i] *Istituto Nazionale di Fisica Nucleare, Sezione di Pavia, 27100 Pavia, Italy*
[j] *Dipartimento di Fisica Nucleare e Teorica, Università di Pavia, 27100 Pavia, Italy*
[k] *Physik-Institut, Zürich University, CH-8057 Zürich, Switzerland*
[l] *Dipartimento di Chimica e Fisica per l'Ingegneria e per i Materiali, Universit`a di Brescia, 25123 Brescia, Italy*

**Abstract.** We demonstrate three-dimensional annihilation imaging of antiprotons trapped in a Penning trap. Exploiting unusual feature of antiparticles, we investigate a previously unexplored regime in particle transport; the proximity of the trap wall. Particle loss on the wall, the final step of radial transport, is observed to be highly non-uniform, both radially and azimuthally. These observations have considerable implications for the production and detection of antihydrogen atoms.

## INTRODUCTION

Imaging techniques have played an important role in trapped particle studies. Dumping particles onto a collimated Faraday cup, or on a screen viewed by a CCD camera, is now a standard technique which gives a $z$-integrated plasma shape [1-3] ($z$ is the direction along the magnetic axis). Detection of laser fluorescence is another common technique for trapped atomic ions where convenient transition lines exist [4,5]. This method can also provide particle velocity from the laser Doppler shift [6].

---

[1] Invited talk at NNP03, Workshop on Non-Neutral Plasmas 2003
[2] Corresponding author: e-mail: Makoto.Fujiwara@cern.ch

Antiparticle annihilation imaging can give information complementary to the above more conventional methods. Depending on the density of the residual gas in the system (see below), antiprotons can annihilate either on gas, or on the trap wall as a result of radial transport. If the vacuum is sufficiently high and the annihilation on the gas is negligible, antiproton imaging is uniquely sensitive to particle losses at the trap wall. It thus allows investigation of particle transport processes in the yet un-explored regime, the proximity of the trap wall.

O'Neil's confinement theorem [7] states that, for an axially symmetric system in a uniform magnetic field, due to the conservation of canonical angular momentum, the mean-square radius of trapped particles is approximately constant, ensuring confinement of non-neutral plasmas. In actual experiments, however, plasmas expand at a finite rate, eventually leading to deconfinement. Starting from the pioneering works of the 1980s [8], radial particle transport across magnetic field lines has been the subject of extensive studies. As such, there is now a large body of evidence which suggests that the radial transport is driven by mechanical and field asymmetries, inherent in all trap constructions [9]. However, while there are notable recent developments [10], the exact mechanism is not yet completely understood. With our imaging technique, we show in this report that the particle loss at the trap wall, the final step of the radial transport, occurs in a manner that is highly non-uniform, both axially and azimuthally.

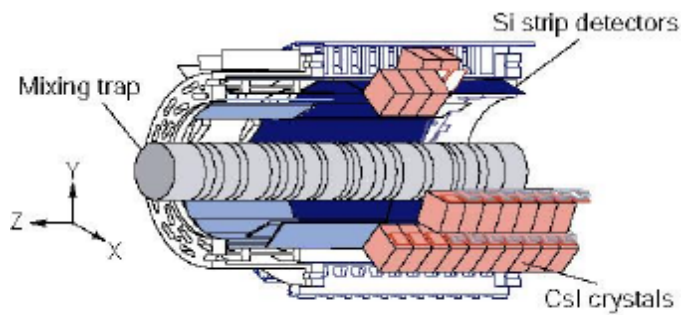

Figure 1. A schematic view of the apparatus. Two layers of double-sided silicon strip detectors and 192 CsI crystals surround the mixing trap. The signal from total of 8192 detector channels are read out via flash analogue-to-digital converters (ADCs) and written onto a disc at a rate of up to 40 Hz.

## EXPERIMENT

The present measurements are performed using the ATHENA apparatus, which recently achieved the first production of cold antihydrogen atoms [11]. Detailed descriptions of ATHENA are given elsewhere [11-15]. The Antiproton Decelerator (AD), located at the European Organization for Nuclear Research (CERN) in Geneva, provides a pulse of 5 MeV antiprotons every 100 s. This is presently the world's only source of low energy antiprotons. The antiprotons are dynamically captured by briefly opening the potential wall at the entrance side. Cold electrons, preloaded in the trap, cool the antiprotons via Coulomb collisions [16]. The antiproton capture and cooling can take place in either of two separate traps; the catching trap, or the mixing trap. In the former case, the cooled antiprotons are adiabatically transferred to the mixing trap after the catching and cooling [17]. The traps (inner radius 1.25 cm) are held at a temperature between 15 and 40 K. The imaging detector is kept at 140 K and housed in a separate vacuum (Fig. 1). Typically, $10^3$ to $10^4$ antiprotons together with $10^7$ to $10^8$ electrons are stored in the mixing trap for these measurements. Antiprotons annihilate either on the residual gas, or if they reach the trap wall, on the surface of gold-plated electrodes.

We observe that keeping the electrons together with the antiprotons shortens the storage time of the latter. This effect, illustrated by the data presented in Fig. 2, is possibly due to an enhancement of the radial transport of the antiprotons due to electron collective effects, and merits further dedicated study. For the purpose of the work here, however, we used this effect simply to accelerate the radial loss. It should be noted, though, that the number of the electrons, or lack thereof, does not affect the main conclusions reported here.

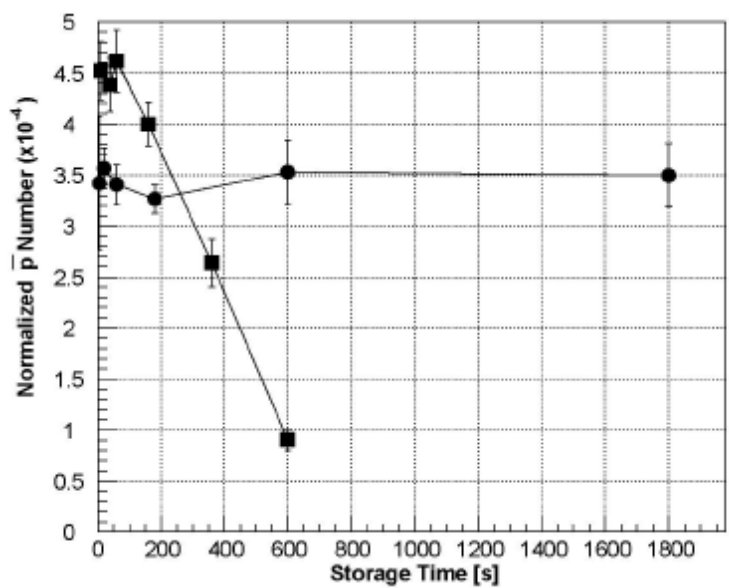

**FIGURE 2.** Number of antiprotons as a function of storage time. The squares refer to measurements with $10^8$ electrons kept in the trap, whilst the circles are with less than $10^5$ electrons remaining. The trapped antiprotons are counted with external annihilation detectors [17], and are normalized to the incoming antiproton number, which in turn is measured by a calibrated beam detector [18].

## ANTIPROTON IMAGES

Antiproton annihilations produce several charged and neutral mesons (mostly pions). The average charge multiplicity depends on the target nucleus; 2.6 for gold and 3.0 for a proton [19]. The charged particles are detected by two layers of double-sided silicon micro-strip detectors. A signal in one of the layers is a "hit", and two hits from two layers are fitted to a straight line to determine a "track". From the intersection of two or more charged tracks, an annihilation "vertex" is determined. A collection of the vertices thus represents the three-dimensional distribution of antiproton positions at the time of their annihilations. Unmeasured curvature of the pion tracks in the 3 T magnetic field is the dominant source of the 4 mm (1$\sigma$) vertex reconstruction uncertainty.

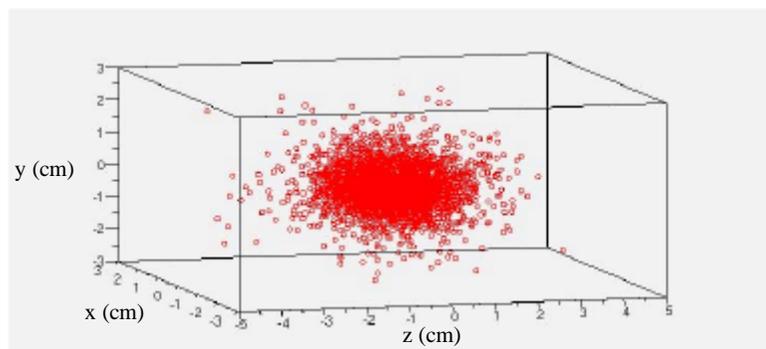

**Figure 3.** Three dimensional imaging of trapped antiprotons

Figures 3 and 4-I (a) show measurements in a harmonic trap with a depth of 30V and length about 5 cm with a relatively high gas pressure in the system, of the order of $10^{-11}$ mbar as estimated from the antiproton lifetime. In these conditions, annihilations on the residual gas (or ions) dominate. Thus, the image obtained corresponds to the distribution of antiprotons in a quadratic potential trap, and is azimuthally symmetric, as expected.

A striking pattern emerges if the residual gas pressure is reduced to below $10^{-13}$ mbar, as shown in Fig. 4-I (b). Annihilations are then observed to be highly non-uniform both azimuthally and axially, and are localized in a few "hot spots". As illustrated in Figs. 4-II and III, the existence of hot spots is a universal feature of charged particle loss in our trap.

Figure 4-II shows images from a series of measurements using only one electrode to create a trap well. A potential of –140 V, with respect to the rest of the grounded electrodes, was applied. The antiproton annihilations take place in the potential well regions, as expected, but they are localized in both $z$ and $\phi$, where $\phi$ is the azimuthal angle of the vertices seen from the trap axis. Figure 4-III illustrates results of measurements with different numbers of electrodes used to form wells. Either –140 V or –50 V was applied, the value of which did not change the features of the images. Again, annihilations are localized. We observe that the annihilation hot spots are clustered near the edge of the electrodes, and their number grows with the number of electrodes used for the well. Azimuthally segmented four-sectored electrodes are seen to enhance annihilations in some (but not all) cases. For example, see the top panel of Fig. 4-III, where the segmented electrodes are depicted as "SE".

(I)

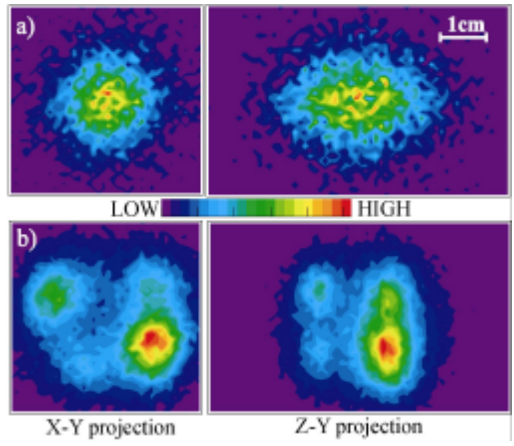

(II)

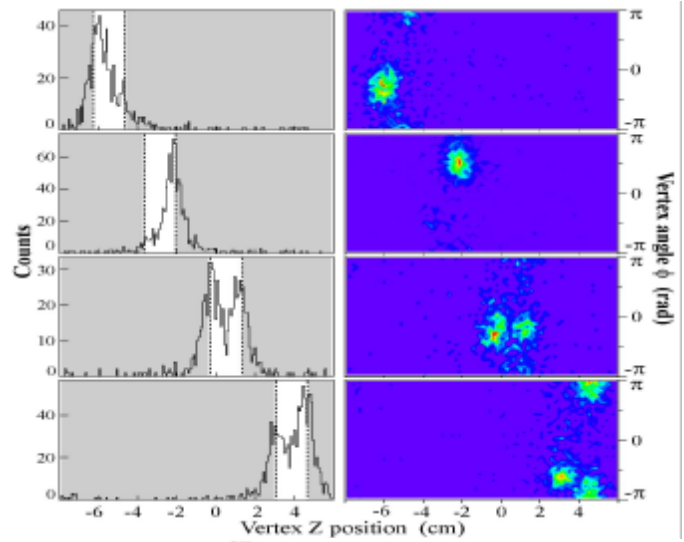

(III)

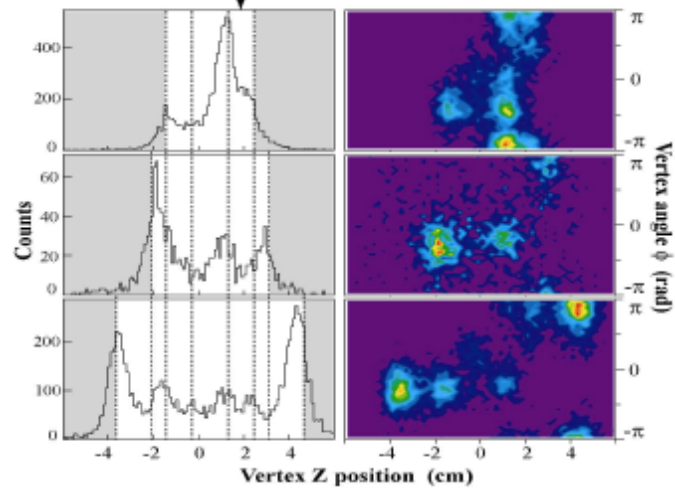

**Figure 4.** Antiproton annihilation images. See text for details.

# MONTE CARLO SIMULATIONS

In order to quantitatively understand the observed images, we have performed detailed Monte Carlo (MC) simulations, based on the GEANT simulation package. In our simulations, antiproton annihilation on protons is assumed, and a tabulated branching ratio is used to generate annihilation products, both charged and neutral. Interactions of these particles with our detector and the apparatus are simulated, including electromagnetic and hadronic cascades in the magnet materials. The apparatus geometry is directly imported from a CAD program, and the measured module-by-module detector efficiencies are included in the calculations. The code generates simulated data in the same structure as the experiment, and the same analysis program is applied to both.

The results of the simulation are compared with the data in Fig. 5. A radial distribution of antiproton annihilations for the high vacuum case (from Fig. 4-I (b)), and a simulated distribution assuming a point annihilation source on the trap wall are plotted in Fig. 5 (a). The good agreement between the experiment and the simulation establishes that most of the annihilations occur on the wall (r=1.25 cm). The structure near the peak of the simulated distribution is due to reconstruction errors caused by the curvature of the charged track. It disappears in simulations with the magnetic field set to zero.

The radial distribution of the measured data for the high density case from Fig. 4-I (a) are also plotted in Fig. 5 (a). As is evident, our imaging can clearly distinguish between the distributions for predominantly gas annihilations and those resulting from wall annihilations.

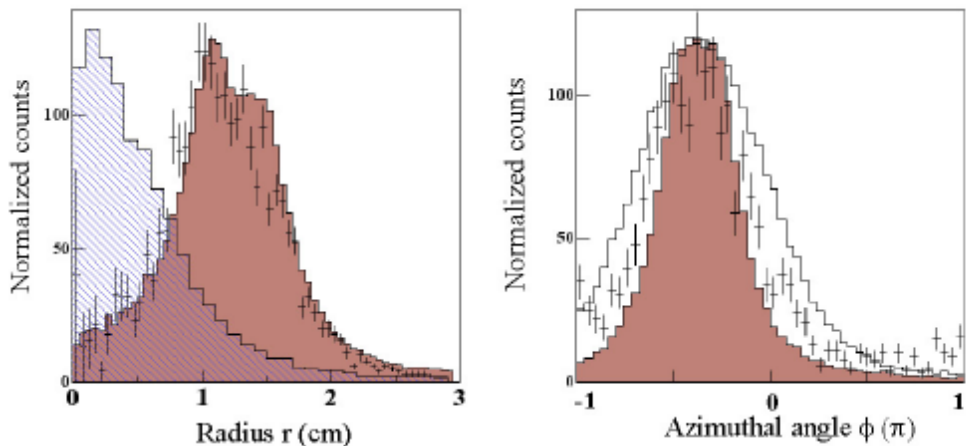

**FIGURE 5.** (a) Comparison of radial ($r$) annihilation distributions ($dN/rdr$) for the data from the measurement at low background gas pressures (error bars) and the MC simulations assuming annihilations on the trap wall (dark histogram). Also shown are the data for the high pressure measurement (diagonally filled histogram). (b) The azimuthal ($\phi$) angular distribution of the annihilation (error bars) and its comparison with the MC assuming point source annihilation (dark histogram). Also shown is a MC assuming an extended annihilation source (±4 mm).

We now focus on one of the hot spots from Fig. 4-I (b), in order to study the extent to which it is localized. Figure 5 (b) is a comparison of the annihilation azimuthal angle distributions for the experimental data and the simulations. Also included are histograms for Monte Carlos assuming both a point source annihilation on the wall, and assuming an extended source spot of ±4 mm. From the comparison, we can exclude a source extent of this size.

## DISCUSSION

While it is clear that the presence of hot spots is a result of asymmetries in the system, the underlying mechanism resulting in loss localization is not completely understood at the present time. The effects due to image charge and surface field irregularities complicate the dynamics of particle transport in the proximity of the trap wall.

Various measurements were performed to establish the universality of the loss localization. We observed the hot spots in all cases when antiprotons annihilate on the wall. This was regardless of the details of the antiproton and electron (re)loading procedure and the values of the potentials. Hot spots are present, even though electrons were removed the electrons from the trap, indicating that the loss localization mechanism is not due to collective plasma effects of the electrons, but is dominated by the single particle transport properties of the antiprotons. Note that the antiproton density of is low ($<10^3$ cm$^{-3}$) in these measurements.

The present observations have major implications for the detection of antihydrogen annihilations. Our initial observation of antihydrogen was based on simultaneous detection of antiproton and positron annihilations at the same place. While antiproton annihilation detection is efficient (the vertices can be reconstructed with about 50% efficiency), positron detection is more difficult due to the low intrinsic efficiency of the CsI crystals, and the presence of background. Thus, the overall efficiency was about 0.2% for fully reconstructed antihydrogen events, where both the charged vertex and back-to-back gamma rays were detected. Our finding that neutral antihydrogen atoms annihilate on the wall in a radially uniform manner [11], whereas charged antiprotons produce hot spots, can provide a new and effective signature of antihydrogen annihilations. An antihydrogen detector without the need to register gamma-rays can be envisioned, a considerable simplification when compared to the present system.

Imaging profiles of antiprotons obtained in high density cases (Fig. 4-I (a)) can provide useful information relevant in our quest to understand the antihydrogen production processes. The image obtained represents the spatial distribution of antiprotons, convoluted with the gas (or ion) distribution on which they annihilate. The observed distributions of the antiproton cloud in Figs. 3 and 4-I (a) have an aspect ratio of approximately 2. This is in contrast to the positron plasma aspect ratio (4 –7), as determined by modes analysis [15] based on Dubin's cold fluid model [20]. If we assume that the effects of residual gas on the particle dynamics is negligible on the time scale of the measurements (a few minutes), the observed difference in the cloud radii may explain the apparent partial mismatch in the radial overlap between antiprotons and positrons, indicated by our measurements of positron cooling of antiprotons [21].

Another application of antiproton imaging is illustrated in Fig. 6. By moving the trap well, and measuring the annihilation positions, we can determine our trap electrode positions, relative to the detector, over a wide range of axial positions as shown in Fig. 6 [22]. Imaging of antiproton annihilations thus permitted detector position calibration at 1 mm precision, a task otherwise nontrivial in the present setup. The precise calibration shown here is an important input to the physics analyses using the ATHENA antihydrogen annihilation detector.

As we have previously stated, our imaging position resolution is, at present, limited by the unmeasured curvature of the charged tracks. This could be improved in a future apparatus by using three or more layers of Si strips. Our detector readout rate (~40 Hz) limits the physical processes that can be imaged to relatively slow ones (such as the radial loss reported here), and much faster processes, e.g. bursts of annihilations due to diocotron instability, cannot be readily imaged. The implementation of the signal level discrimination at the ADC level (so called zero suppression) is in progress and can improve the readout rate by up to a factor of 10.

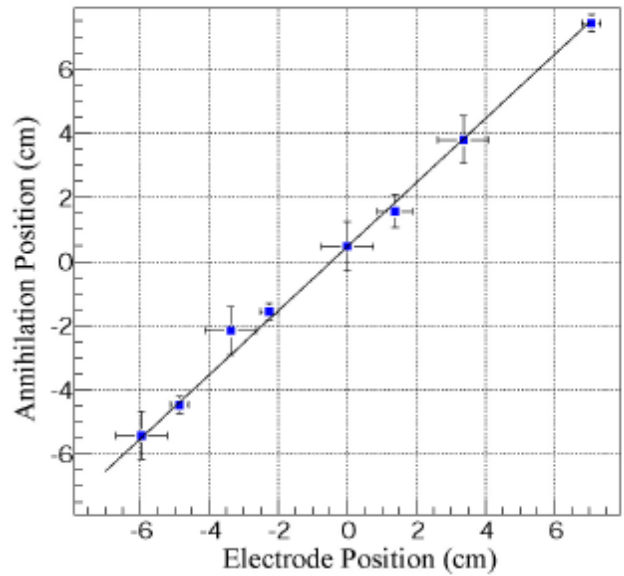

**Figure 6**. The correlation between the trap well positions and the measured annihilation positions.

## CONCLUSIONS

In this paper, we reported imaging of antiproton distributions via reconstruction of annihilation vertices. With this new technique, we probed the previously unexplored final step in the radial transport of trapped charged particles; a regime in the proximity of the trap wall. We observed that antiproton annihilations on the trap wall are localized in all cases, an effect which may be applicable to other Penning (and related) systems. Several implications for antihydrogen production and detection were discussed.

The main disadvantage of the antiproton imaging technique for trapped particle studies is the scarcity of antiprotons. In the near future, we will extend our antiparticle imaging studies by using positron annihilations, much like positron emission tomography in medical applications.

## ACKNOWLEDGMENTS

We gratefully thank CERN's AD crew and J. Rochet for providing essential support, Professors J. Fajans, H. Higaki, A. Mohri, and Y. Yamazaki for valuable discussions, and A. Cavanagh for a critical reading of the manuscript. This work was supported in part by MEXT and RIKEN (Japan), CNPq (Brazil), SNF (Denmark), INFN (Italy), SNF (Switzerland), and the EPSRC (UK).

## REFERENCES

1. Dubin, D. H., and O'Neil T. M., Rev. Mod. Phys. **71**, 87 (1999).
2. Davidson, R. C., *Physics of Nonneutral Plasmas*, (Imperial College Press, London, 2001)
3. *Non-Neutral Plasma Physics*, edited by Anderegg, F., et al. Vol. 4 (AIP Conference Proceedings, New York, 2002).
4. Larson, D. J., et al., Phys. Rev. Lett. **57**, 70 (1986).
5. Brewer, L. R., et al., Phys. Rev. A **38**, 859 (1988).
6. Mitchell, T. B., et al., Opt. Express **2**, 314 (1998).
7. O'Neil, T. M., Phys. Fluids **23**, 2216 (1980).
8. Malmberg, J. H., and Driscoll, C. F., Phys. Rev. Lett. **44**, 654 (1980); Driscoll, C. F. and Malmberg, J. H., ibid. **50**, 167 (1983); Driscoll, C. F., et al., Phys. Fluids **29**, 2015 (1986).
9. Notte, J., and Fajans, J., Phys. Plasmas **1**, 1123 (1994); Eggleston, D. L., and O'Neil, T. M., ibid. **6**, 2699 (1999); Chao, E. H., et al., ibid. **7**, 831 (2000); Kriesel, J. M., and Driscoll, C. F., Phys. Rev. Lett. **85**, 2510 (2000); Kabantsev, A. A., et al., ibid. **87**, 225002 (2001); Sarid, E., et al., ibid. **89**, 105002 (2002).
10. Kabantsev, A. A., and Driscoll, C. F., Phys. Rev. Lett. **89**, 245001 (2002); Kabantsev, A. A., et al., Phys. Plasmas **10**, 1628 (2003).
11. Amoretti, M., et al., Nature 419, 456 (2002).
12. Amoretti, M., et al. submitted to Nucl. Instr. Meth. A., 2003
13. Regenfus, C., Nucl. Instrum. Methods A **501**, 65 (2003).
14. Fujiwara, M. C., et al., Nucl. Instrum. Methods B. in press (e-print archive: hep-ex/0306023).
15. Amoretti, M., et al., Phys. Plasmas **10**, 3056 (2003); Phys. Rev. Lett. **91**, 055001 (2003).
16. Gabrielse, G., et al., Phys. Rev. Lett. **57**, 2504 (1986); **63**, 1360 (1989).
17. Fujiwara, M. C., et al., Hyperfine Interact. **138**, 153 (2001).
18. Fujiwara, M. C., and Marchesotti, M., Nucl. Instrum. Methods A **484**, 162 (2002).
19. Bendiscioli, G., and Kharzeev, D., Rivista Nuovo. Cim. 17(6), 1 (1994).
20. Dubin, D. H., Phys. Rev. Lett. **66**, 2076 (1991).
21. Amoretti, M., et al., to be published.
22. Bouchta, A., and Fujiwara, M. C., ATHENA Technical report (2002).